# Further detailing of the Bose-Einstein negative-*U* state in the high temperature superconducting cuprates.

**John A. Wilson**

H.H.Wills Physics Laboratory
University of Bristol
Tyndall Avenue
Bristol BS8 1TL
U.K.

**Abstract**

The cause of the sharp and universal optimization of the HTSC condensation energy at the hole doping concentration of *p* = 0.19 is identified within the context of the boson-fermion negative-*U* modelling and stripe phase electronic organization.

Some recent structural, optical and positron annihilation experiments adding further support to this type of modelling of HTSC are briefly examined.

e-mail; john.a.wilson@bris.ac.uk

___

The work of Loram, Tallon and others [1], especially the specific heat analysis, indicates that the really special hole concentration within the superconducting dome of the HTSC cuprates is not the point where $T_c$ is brought to a maximum ($p$ = 0.16), but the somewhat increased concentration of $p$ = 0.19 where the superconducting condensation energy per pair comes rather sharply to a maximum. While contemplating appropriate cluster geometries with which to proceed towards a dynamical mean-field theory (DMFT) cluster analysis for the negative-$U$ state long advocated in [2], it quickly became apparent that the most advantageous form, as expressed within a stripe phase framework, would be that given in figure 1. This takes the charge-alternating form of the hole segregating stripes, with charge transfer permitted in the key, high scattering, antinodal directions, $k_x$ and $k_y$, the stripes at this concentration being placed in 'crossed' geometry. The latter charge pairing pathways which supply the vital negative-$U$ double loading are favoured when at right angles to each other for reasons outlined in [3]. The achieved long-lived local-pair fluctuations at the central trivalent site (associated in $La_{2-x}Sr_xCuO_4$ (LSCO) with a Sr substituent) are written in the notation introduced in [2a] as $^{10}Cu_{III}^{2-}$. Here III denotes the formal valence of the site, 10 shows the site's temporary outer electron count, and 2- marks the instantaneous departure in this cell charge loading from local charge neutrality. Taking the above cluster and repeating it as densely as possible for maximum efficacy generates the square $4a_o$ x $4a_o$ structure appearing in figure 2. And what is the hole content of this supercell? It is 3, yielding the crucial $p$ value of $3/16$ or 0.1875.

The stripe structures that originally I contemplated in [4] were of the bond-centred type. This automatically generates a supercell repeat dimension twice that for the 'thin-walled' atom-centred variety above. For the 'thick-walled' option the charge periodicity becomes identical to the magnetic superperiod. The latter was first recorded for LSCO and LBCO via inelastic neutron spin scattering. The stripe ordering is in general only dynamically organized except near $p = 1/8$ [5]. A bond-centred geometry was embraced in [4] for the optimally and underdoped samples examined originally since it is compliant with RVB spin quenching. The latter was perceived as responsible for the observed low magnetic susceptibility, spin-gapped condition [6], the structure interior to the stripe domains being viewed as built upon $d^9$ 4-spin square plaquets. When the hole content in the system is raised beyond $p = 1/8$, the stripes rapidly become highly compacted, and it would appear advantageous then for the atom-centred stripe to be adopted. The latter condition has the potentially deleterious effect of restricting spin plaquet formation. In fact the low temperature susceptibility of all HTSC cuprate systems is just starting to climb back up again by the hole doping concentration for which $T_c$ maximizes [7]. In figure 2 one notes that the interior content of each stripe domain now amounts to just 9 sites. Being an odd number this implies spin quenching cannot be expected to be complete. In a regular array of stripe domains successive domains would need to be coupled in antiphase across the stripes if full spin gapping were to be maintained. It will be very interesting to find what μSR makes of the low temperature magnetic condition in a $p = 0.19$ sample. The charge fluctuation currents in this broken symmetry, more magnetic circumstance might well be source to the optical dichroism being reported still with HTSC materials [8]. Broken time reversal symmetry dichroism effects first were considered by Simon and Varma [9], and subsequently by Chakrvarty *et al* [10]; however the simple current paths suggested there did not look appropriate.

When the structure shown in the present figure 2 is globally subjected to the form of fluctuation represented in figure 1, one attains the situation appearing in figure 3. Portrayed here in static, quasi-classical, mean-field manner is the charge pattern acquired under the frozen fluctuation. One sees it to involve a $2a_o$ charge transfer wave within the stripes, negative charge being moved into all the hole sites. Transfer into the central stripe intersection sites will in practice be most marked wherever those sites locally are stabilized strongly towards $Cu_{III}$ (*i.e.* in LSCO by the presence of an adjacent Sr substituent atom). Since such substituent ions are weakly screened within the Mott insulating host at the high growth temperatures, they are likely to be maximally dispersed, and hence frequently encountered. (See figure 3 in [4] to get a feel for the situation at these contents). What happens at the corner of the stripe domain, especially wherever centred upon such a substituent, is to be gathered from figure 3. A five-cell cluster is formed over which, provided the fluctuation lives long enough, the lattice is able to retreat annularly around the expanding central cell taking on the charge double loading. This accommodating contraction is aided by the fact that the $d_{x^2-y^2}$ electrons being transferred had been strongly antibonding. (N.B. the cell volume of $La_2CuO_4$ is 95.7 Å$^3$ as compared with only 94.7 Å$^3$ for $La_2NiO_4$). However what is experienced at the central $d^{10}$ site itself in the charge fluctuation is not so much a lattice controlled effect as one of total electronic reorganization of *all* the valence states. All bonding/antibonding between the Cu d shell and O p shell states is terminated with the shell filling that the above fluctuation entails. This is what brings negative-*U* character to the doubly-loaded, $d_{x^2-y^2}$ symmetry state in $^{10}Cu_{III}^{2-}$ [3, figs.1 and 5]. For the case of divalent AgO, a static $d^{10}$ loading imparts non-metallic behaviour there in disproportionated fashion, but for the mixed-valent doped cuprates this more stable configuration is gained locally without sacrificing overall metallicity, and in due course permitting superconductivity to issue from the $d_{x^2-y^2}$ pairings. The negative-*U* states by chance fall in near-resonance with $E_F$ [3,2]. Conduction in these systems, not too far removed from the Mott transition, is controlled by the mixed-valent intersite charge process -

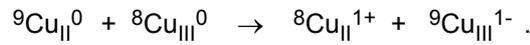

$$^{9}Cu_{II}^{0} + {}^{8}Cu_{III}^{0} \rightarrow {}^{8}Cu_{II}^{1+} + {}^{9}Cu_{III}^{1-} .$$

Under the global change incurred between figures 2 and 3 two such transfers per domain occur in conjunction with the double-loading negative-*U* fluctuation -

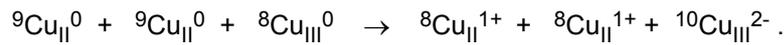

$$^{9}Cu_{II}^{0} + {}^{9}Cu_{II}^{0} + {}^{8}Cu_{III}^{0} \rightarrow {}^{8}Cu_{II}^{1+} + {}^{8}Cu_{II}^{1+} + {}^{10}Cu_{III}^{2-} .$$

The $4a_o$ period to the above dynamic charge transfer pattern, and accordingly to the coupled lattice relaxation, is reflected in the marked phonon mode softening of the basal longitudinal optic mode found halfway down the $k_x$ and $k_y$ axes. Such charge-lattice intermode coupling was discussed at length in [2b] in connection with Chung *et al*'s neutron scattering data [11a]. Recent comparable data have just appeared from Pintschovius *et al* [11b] using detwinned samples which help to clarify and amplify certain points. Reasons are advanced in [2b §6] for the detected instability onsetting close to $\xi_{opt}(0)$ (~ 16Å), *i.e.* $4a_o$, within the context of the boson-fermion negative-*U* modelling given there.

    Local lattice responses which develop in the vicinity of $T_c$ have long been in evidence from the EXAFS data [12], but these are not sufficiently discriminatory to permit more than a bipartite analysis, and therefore cannot closely probe the negative-*U* fluctuation. Since the local pair centre is doubly charged it is possible however to couple directly to this feature in positron annihilation work. Fairly recently a Doppler broadening line-width study (S parameter) on *single crystal* $Bi_2Sr_2CaCu_2O_{8+\delta}$ indicated a non-structural event some 20 degrees above $T_c$ in which the number of quasiparticles becomes reduced [13]. One ought to attempt to reproduce this important result on crystals for other appropriate HTSC systems.

Certain optical and laser techniques are able to supply information directly about the local pair population [see 3,14]. One type of optical experiment much discussed at the moment, namely I.R. spectral weight shift, provides further clear indication of the high energy electronic nature of the HTSC mechanism, changes in reflectivity across $T_c$ running up at least to 1.0 eV. It has to be recalled that although the superconducting gap energy has a scaling $2\Delta \sim 5kT_c$ (*i.e.* $\sim$ 40 meV or 320 cm$^{-1}$ in BSCCO), the local pair metastability energy (re $^9$Cu$_{III}^{1-}$) is set by $|-U_{eff}/2|$ per electron. Ref.[3] identifies this latter energy as 1.5 eV. It is very interesting to find Boris *et al* [15] in their very accurate spectral weight work and analysis indicate that as $T$ falls through the region of $T_c$ the spectral weight also falls (despite the claims of others to the contrary). This implies a rise in the kinetic energy (as customary). However because of the way in which the observed high energy modifications (beyond 0.3 eV) cause the plasma frequency to become depressed, the latter shows that the measured spectral weight decrease is to be associated with a *fall* in carrier density, rather than some upward renormalization in m* (or any association with the spin pseudogap). Besides tracking reduction in the quasiparticle density, the analysis demonstrates in addition, via its spectral broadening parameter $\Delta\gamma(\omega)$, that intense scattering is endemic in the system, and that this bears extrapolation to 1.5 eV – our local-pair dissolution energy. Such portentious results clearly demand much closer attention.

The author would like to acknowledge The Leverhulme Trust for their financial support through an Emeritus Research Fellowship.

**Figure captions**

**Figure 1**. The situation at the intersection of two perpendicular charge-alternating stripes, mutually phased such that the intersection cell is a $^8Cu_{III}^0$. This centre may be doubly loaded to $^{10}Cu_{III}^{2-}$ in the negative-$U$ charge transfer process, particularly if the intersection point is closely associated with a dopant atom (Sr in LSCO). [corner-sharing square basal sections of coordination unit octahedra: shaded squares (red in fig.2), $d^9$ $Cu_{II}$; unshaded (green in fig.2), $d^8$ $Cu_{III}$].

**Figure 2**. 2-D repetition of figure 1 building into a regularized 4-by-4 cell superlattice. The substitution level corresponding to each such domain is 3 (considering the monovalent doping regime of LSCO) and corresponds to the hole density $p$ = 3/16 or 0.1875, under which the HTSC condensation energy is known to become sharply maximized. The classical intersite charge transfer wave along a stripe corresponding to the proposed negative-$U$ charge fluctuation process is the $2a_o$ wave shown at the bottom. The u's and d's indicate the potential spin arrangement for this regularized geometry. Note alternate domains bear here alternating net spin directions for successive 3-by-3 interior arrays of $^9Cu_{II}^0$. [Dark and light green squares, $^8Cu_{III}^0$; red and pink squares, $^9Cu_{II}^0$].

**Figure 3**. The stripe patterning of figure 2 after subjected to a global frozen negative-$U$ charge transfer fluctuation giving $^{10}Cu_{III}^{2-}$ at the cell corners. [Blue squares, $^{10}Cu_{III}^{2-}$; orange squares, $^9Cu_{III}^{1-}$; yellow squares, $^8Cu_{II}^{1+}$; pink squares, unchanging non-stripe $^9Cu_{II}^0$].
Note the process for each 4-by-4 domain involves

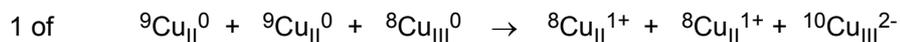

1 of    $^9Cu_{II}^0$ + $^9Cu_{II}^0$ + $^8Cu_{III}^0$ → $^8Cu_{II}^{1+}$ + $^8Cu_{II}^{1+}$ + $^{10}Cu_{III}^{2-}$

the double-loading negative-$U$ local excursion,

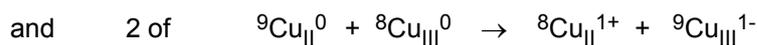

and    2 of    $^9Cu_{II}^0$ + $^8Cu_{III}^0$ → $^8Cu_{II}^{1+}$ + $^9Cu_{III}^{1-}$

the standard mixed-valent intersite conduction process.

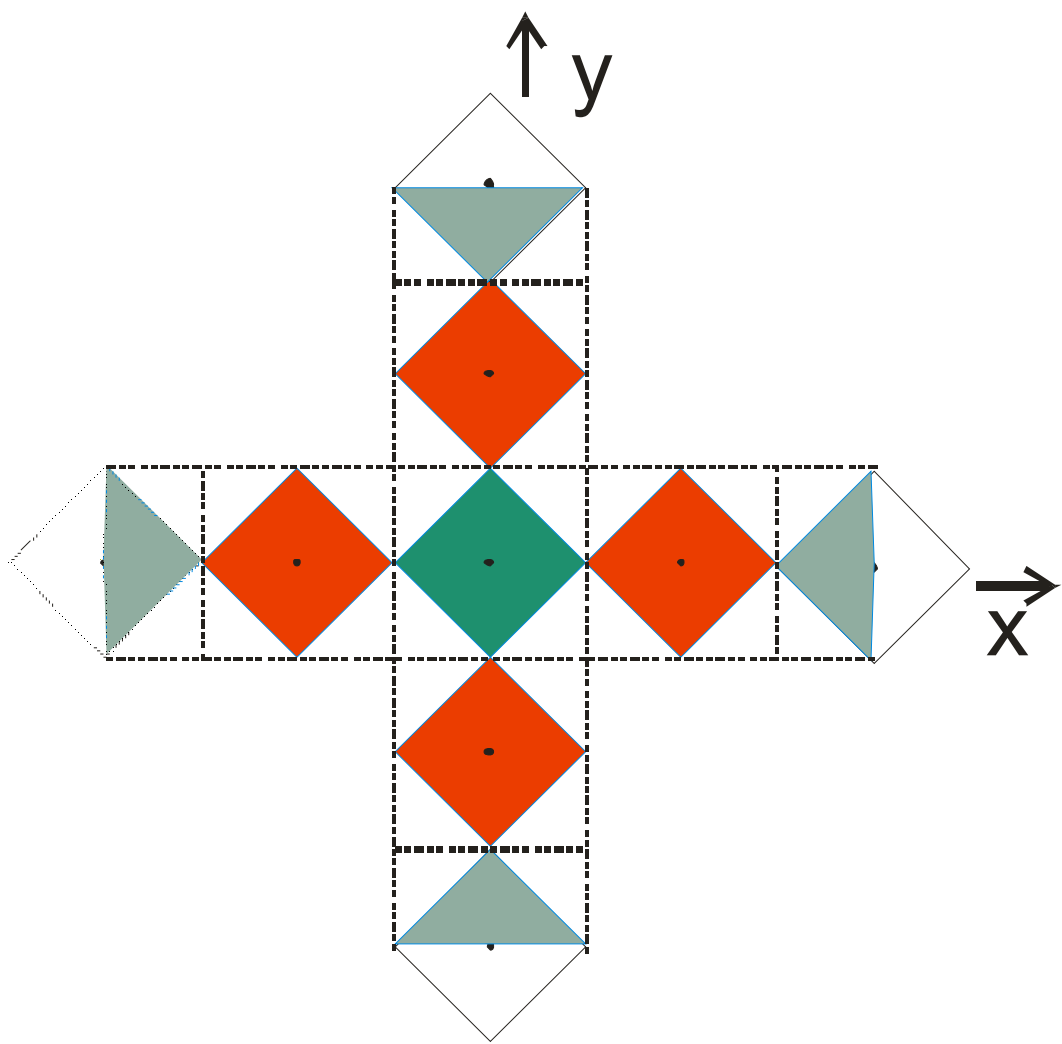

Fig 1 P251

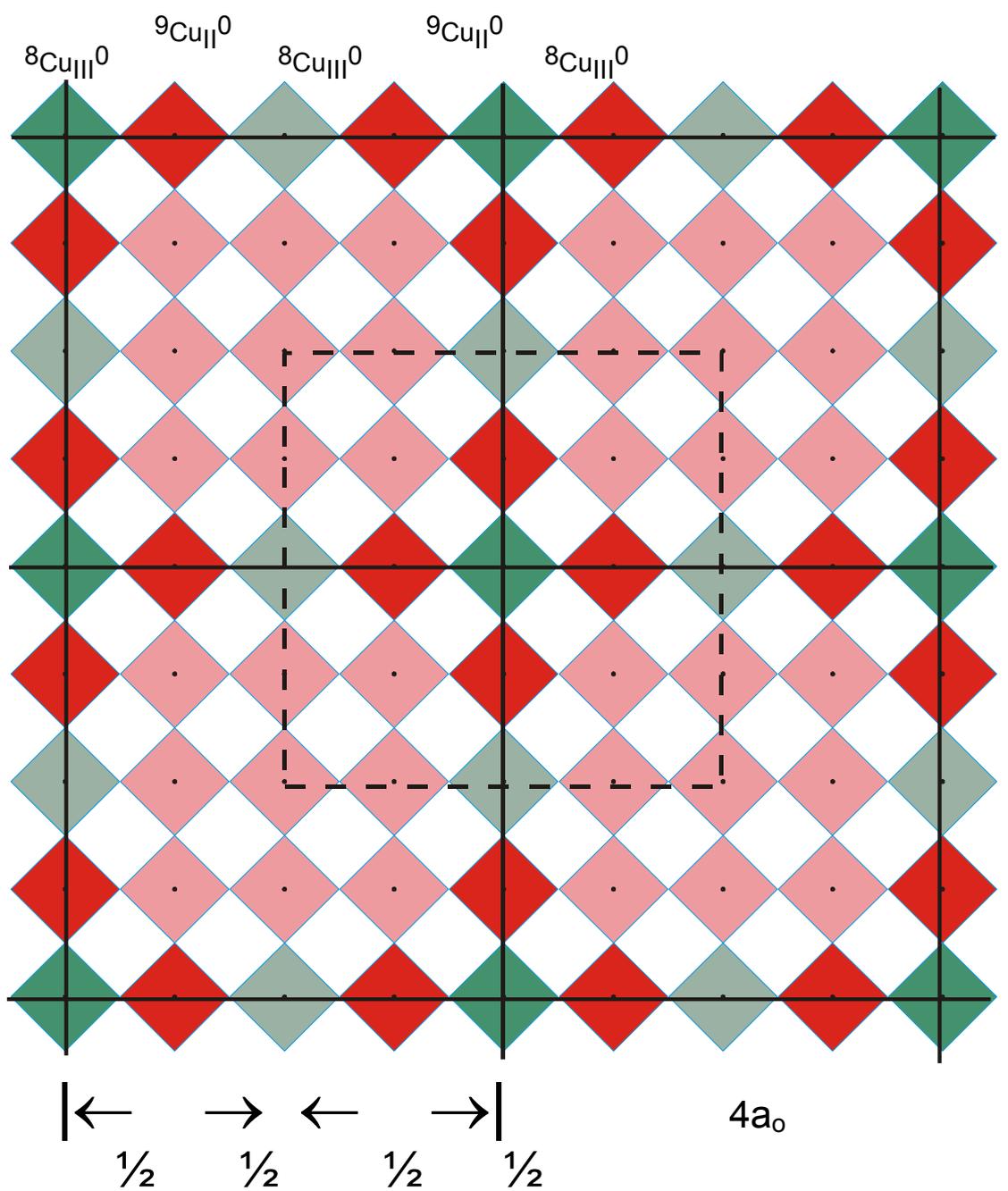

Fig 2 P251

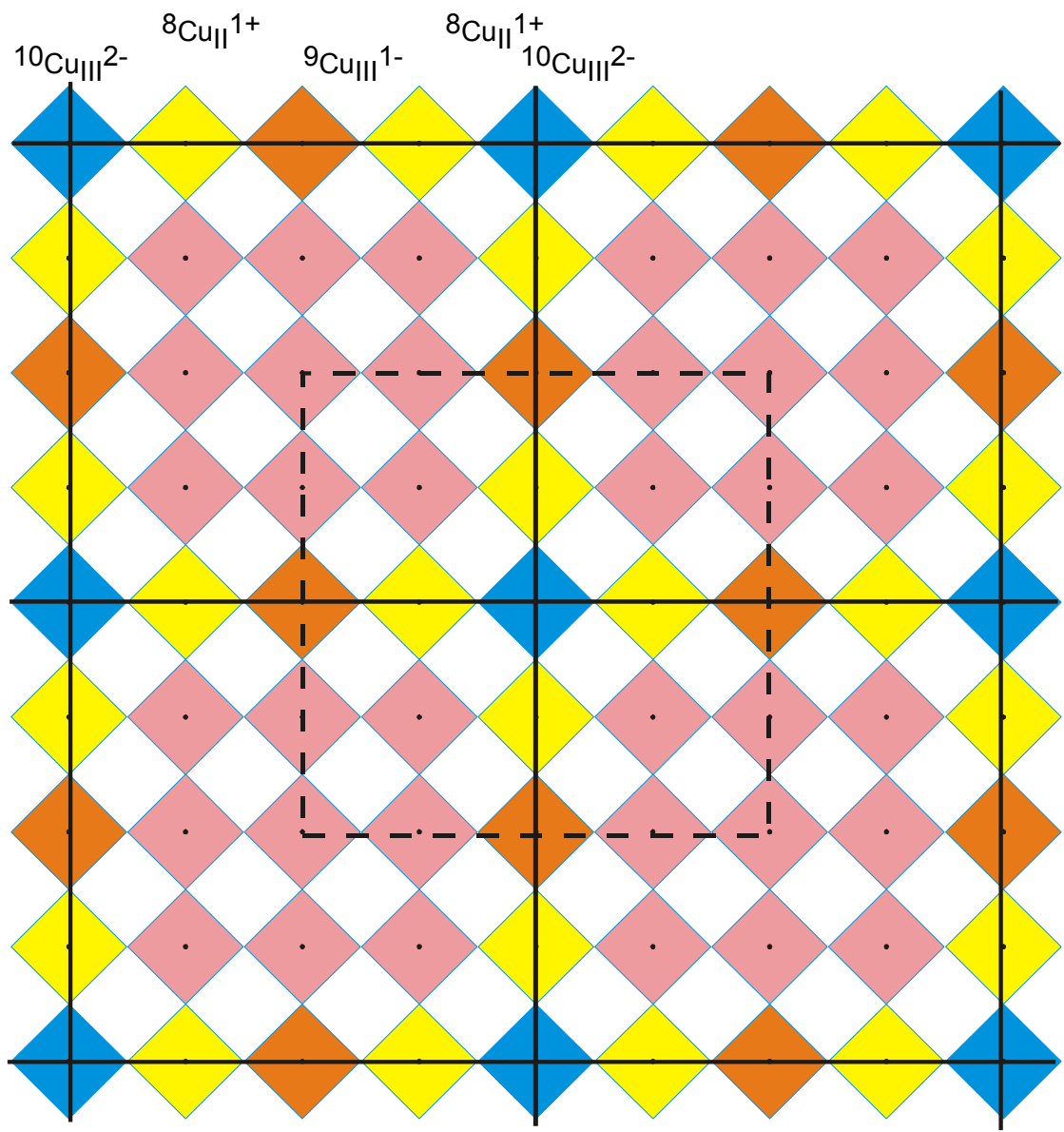

Fig.3 P251